\documentclass[lettersize,journal]{IEEEtran}
\usepackage{amsmath,amsfonts}
\usepackage{algorithmic}
\usepackage{algorithm}
\usepackage{array}
\usepackage[caption=false,font=normalsize,labelfont=sf,textfont=sf]{subfig}
\usepackage{textcomp}
\usepackage{stfloats}
\usepackage{url}
\usepackage{verbatim}
\usepackage{graphicx}
\usepackage{cite}
\begin{document}

\title{Simulation analysis of a direct/indirect conversion dual-layer X-ray detector for early identification of coronary artery calcium}

\author{Akyl Swaby, Linxi Shi, Kaitlin Hellier, Adam S. Wang, and Shiva Abbaszadeh
\thanks{A. Swaby, K. Hellier, and S. Abbaszadeh (Email: sabbasza@ucsc.edu) are with the Department of Electrical and Computer Engineering, University of California Santa Cruz, Santa Cruz, CA 95064, USA.}
\thanks{L. Shi and A. S. Wang are with Department of Radiology, Stanford University, Stanford, CA 94305, USA.}}



\maketitle

\begin{abstract}
Coronary artery calcification (CAC) is a strong predictor of future cardiovascular events, and CAC burden is commonly used to stratify patient cardiovascular risk. Although CAC is typically quantified using computed tomography, routine chest radiography offers an opportunity for lower-dose, widely available opportunistic screening if calcifications can be visualized with sufficient contrast. Single-exposure dual-energy chest radiography using multilayer flat-panel detectors can generate low- and high-energy images without the motion artifacts associated with double-exposure techniques and provide material decomposition parameters, including calcification density. 
In this work, we investigate a direct/indirect dual-layer detector consisting of an amorphous selenium (a-Se) direct-conversion top layer and a cesium iodide scintillator-based indirect-conversion bottom layer for enhanced visualization of small CAC. A cascaded linear systems model (CLSM) was used to simulate task-based detectability for a 1 mm calcification embedded in soft tissue. Anatomical noise was incorporated through fitted noise parameters derived from simulated dual-energy lateral chest images. A complementary image-domain contrast-to-noise ratio (CNR) analysis was performed using CT-derived patient anatomy with inserted 1 mm hydroxyapatite calcifications to evaluate CNR as a function of a-Se thickness and dual-energy weighting factor. The CLSM analysis revealed peak task-based detectability at an a-Se thickness of 500–550 \textmu m and a tissue-cancellation weighting factor of $w_s = 1.08$. Detectability remained within a few percent of this maximum across a broad range of a-Se thicknesses (approximately 300–1000 \textmu m). The image-domain CNR analysis yielded maximum CNR at 300 \textmu m a-Se and $w_s=0.93$, with a smaller range of 200-500 \textmu m giving acceptable CNR. These results indicate that near-equal weighting of the two detector layers improves anatomical-noise suppression and CAC conspicuity, while a 300 \textmu m a-Se top layer may provide a practical design compromise between detectability, operating voltage, and manufacturability.
\end{abstract}

\begin{IEEEkeywords}
Amorphous selenium, cascaded linear system model, coronary artery calcium, dual-energy, dual-layer X-ray detector.
\end{IEEEkeywords}

\section{Introduction}
\IEEEPARstart{C}{hest} X-rays (CXR) are among the most common medical imaging procedures and are widely available, including in low-resource settings, while providing low radiation dose and high patient throughput. X-ray flat-panel detectors (FPDs) are important diagnostic tools used to facilitate CXR imaging \cite{spahn2003flat, kalender2003use, kalender2007flat}. As a single layer, FPDs are capable of dual-energy (DE) imaging with a double-exposure \cite{gupta2021dual, fredenberg2018spectral}, differentiating soft tissue from other endogenous or exogenous materials such as bones, contrast agents, and calcium plaques \cite{zhou2018visualization, fischbach2003dual, gilkeson2004digital, kuhlman2006dual, forghani2017dual}. Multilayer FPDs provide enhanced DE imaging by acquiring low-energy (LE) and high-energy (HE) X-ray images in a single exposure (SE). The LE X-rays are absorbed by the top detector, while the remaining HE X-rays can be hardened by an inter-detector filter to increase energy differentiation before being absorbed by the bottom detector \cite{szucs2008single, maurino2019novel}. There has been interest in the development of dual-layer (DL) and multilayer FPDs as a means of acquiring high-resolution images while taking advantage of spectral separation, the absence of motion artifacts, and reduced radiation dose \cite{barnes1985detector, wang2021high,zhao2021effects, ma2020high, kim2018development, mccollough2015dual}. 

Enhanced material decomposition during routine CXR has the potential to assist radiologists in the timely identification of coronary artery calcium (CAC), mitigating cardiovascular-related diseases and improving early detection for asymptomatic patients \cite{willemink2019coronary,hsieh2022estimating}.  CAC scoring is used clinically to classify patients into cardiac risk categories which range from low risk with necessary preventive care, and to high risk of future cardiac events. Calcification sizes, in the early risk category, range from 0.5 \textmu m to 15 \textmu m and, in chronic cases, can grow above 3 mm sizes \cite{mori2018coronary}. Although CAC quantification is typically calculated from computed tomography (CT); however, CT-based procedures are not widely used for population screening due to their cost and radiation burden. DE CXR has demonstrated the ability to visualize CAC; however, conventional double-exposure techniques are susceptible to cardiac and respiratory motion artifacts that limit reliable detection and quantification \cite{boswell2022opportunistic,boswell2025coronary}. Recently, a SE multilayer FPD improved reader confidence for CAC detection compared with conventional composite chest radiographs, demonstrating the promise of multilayer detector technology for opportunistic CAC assessment \cite{rogalla2025opportunistic}.

Table 1 summarizes current efforts in the research and development of multilayer X-ray detectors. The drawback of scintillator-based (e.g., CsI:Tl, GOS, etc.) indirect conversion X-ray detectors is the trade-off between spatial resolution and noise. Thicker scintillator layers improve X-ray stopping power but also increase optical photon spread, which reduces spatial resolution and degrades image quality \cite{badano2002depth, badano2003optical}. In comparison, amorphous selenium (a-Se) based direct-conversion X-ray detectors are widely used in commercial digital mammography because they offer better spatial resolution, a high X-ray absorption coefficient for medically relevant energies (15--25 keV)), good charge transport properties, and low dark current \cite{zentai2009photoconductor, zhao2005amorphous, abbaszadeh2013microfabricated, arnab2017impact}. In a-Se, since there is no optical spread, high spatial resolution can be achieved as long as the layer thickness and applied electric field support efficient charge collection. Considering the good efficiency of selenium for low-energy X-rays, we are investigating a DL configuration consisting of a-Se direct conversion top layer and CsI:Tl indirect conversion bottom layer. While the bottom indirect layer does not fully overcome the resolution trade-off, it compensates for the limited thickness of selenium, which cannot be made very thick due to field and transport constraints, thus enabling better absorption of high-energy photons.  

In this work, we use an integrated simulation framework with two complementary analysis endpoints to optimize the a-Se top-layer thickness and DE weighting factor for CAC assessment.
First, detector metrics derived from a cascaded linear systems model (CLSM) were combined with anatomical noise parameters extracted from patient images within a generalized task-based framework. This allowed us to compute ideal-observer detectability for a 1 mm CAC detection task as a function of a-Se thickness and tissue-cancellation weighting factor, following methods previously reported \cite{richard2005generalized, richard2007optimization, vennart1997icru, richard2008comparison}. 
Second, the same detector metrics were used to generate CT-derived DE-CXR images with 1 mm calcifications to evaluate image-domain CAC conspicuity using contrast-to-noise ratio (CNR). Together, these studies identify both the physics-driven optimum and the practical design range for a direct/indirect DL detector intended for opportunistic CAC visualization in single-exposure DE chest radiography.

\begin{table*}[hbt!]
\small
\renewcommand{\arraystretch}{1}
\caption{Current State-of-the-Art Multilayer Flat Panel X-ray Detectors.}
\centering
\begin{tabular}{ |p{3.5cm}|c|c|c|}
 \hline
 \textbf{Group}& \textbf{Conversion type}& \textbf{Pixel aperture}& \textbf{Detector layer}\\
  & \textbf{\& Material} & \textbf{(\textmu m)} & \textbf{thickness (\textmu m)} \\
 \hline
 Johns Hopkins \cite{wang2021high} &Indirect CsI:Tl &150  &200/550\\
 \hline
 Stanford \cite{shi2020characterization} &Indirect CsI:Tl &150  &200/550\\
 \hline
 KA Imaging \cite{Reveal} &Indirect CsI:Tl &140  &200/400/550\\
 \hline
 Stony Brook \cite{duan2022evaluation} &Direct a-Se/Indirect CsI:Tl &85/85 &200/400\\
 \hline
 UCSC \cite{swaby2022dual} &Direct a-Se/Indirect CsI:Tl &85/85 &*/550\\
 \hline
\end{tabular}

\vspace{1mm}
\textit{* UCSC's top layer thickness represents a topic of this study.}
\end{table*}

\section{Methods}

\subsection{Cascaded Model Simulation}
A CLSM is a useful tool to model the performance of X-ray FPD imaging systems \cite{siewerdsen1998signal, kabir2011modelling, zhao2004indirect, zhao2001effects, richard2005generalized, zhao2003imaging, cunningham2000applied, pacheco2024casymir} as a function of energy (keV) and spatial frequency (mm$^{-1}$). As shown in Fig. 1, the simulation models the imaging chain of detectors (direct or indirect conversion) as a series of independent physical processes, such as signal attenuation, gain, and optical blurring, resulting in the modulation transfer function (MTF), noise power spectrum (NPS), and detective quantum efficiency (DQE)  \cite{kabir2011modelling, pacheco2024casymir}. Noise contributions include the electronic noise from the readout circuit, anatomical noise of overlying structures, and quantum noise from the statistical fluctuations in incident spectrum interaction. Inclusion of anatomical noise yields the ``generalized'' DQE (GDQE) and noise equivalent quanta (GNEQ).

\begin{figure}
\centering
\includegraphics[width=0.49\textwidth]{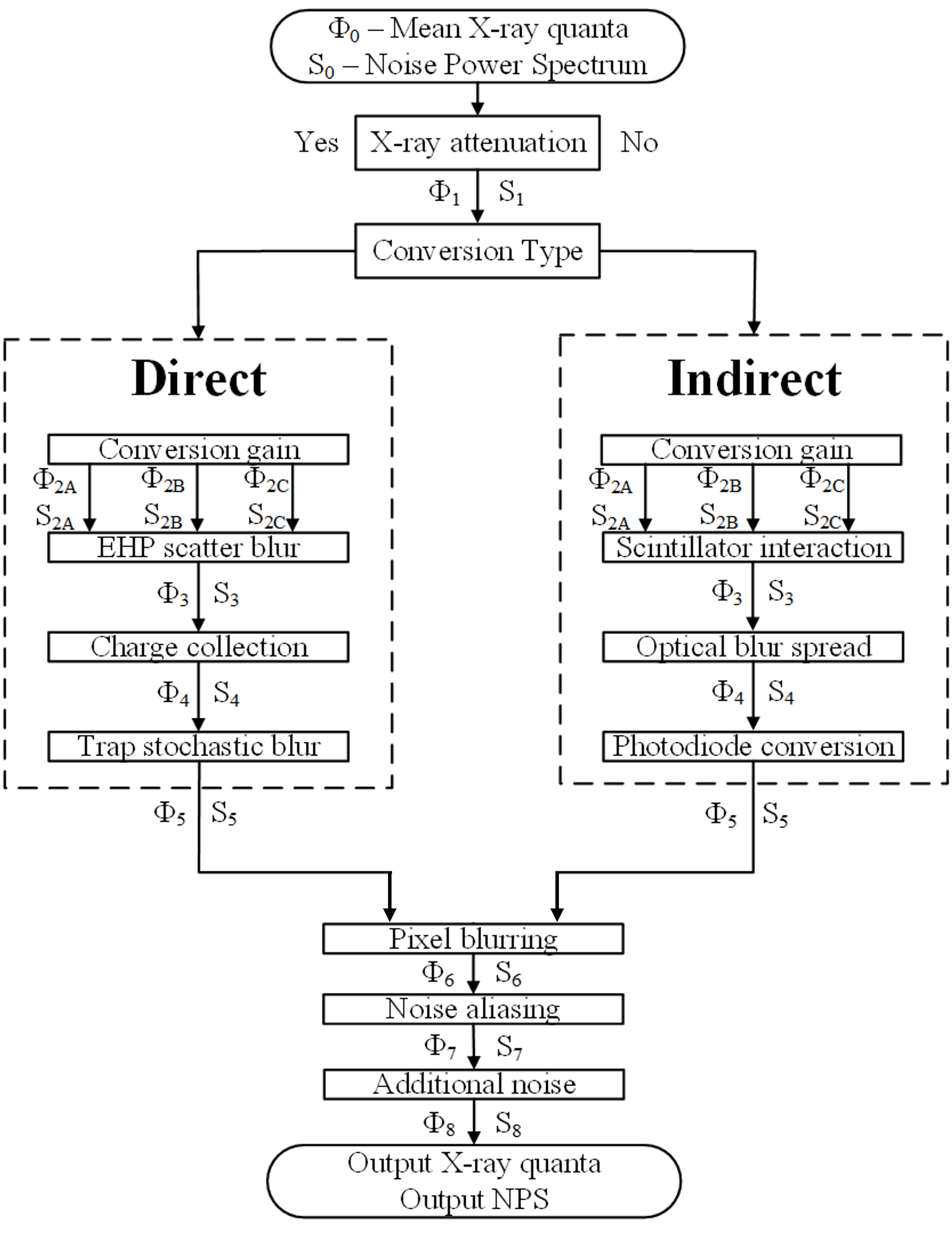}
\label{Cascade}
\caption{Flow chart of the CLSM showing the propagation of
the signal ($\phi$) and noise (S). The two parallel branches show the processes for two detector conversion types: direct and indirect. Once the charge conversion  has been completed, the remaining processes return to follow the same steps.}
\end{figure}

For direct conversion, the CLSM is organized in the following stages \cite{kabir2011modelling}:
(1) X-ray attenuation,
(2) Conversion gain,
(3) Stochastic scattering by the photoconductor,
(4) Charge collection,
(5) Blurring by charge collection,
(6) Deterministic blurring by the pixel aperture,
(7) Noise aliasing, and
(8) Addition of electronic noise.

For indirect conversion \cite{zhao2004x, thacker2004monte, badano2003optical}:
(1) X-ray attenuation,
(2) Conversion gain,
(3) Scintillator interaction,
(4) Photon blur spread,
(5) Photodiode conversion,
(6) Deterministic blurring by the pixel aperture,
(7) Noise aliasing, and
(8) Addition of electronic noise.

\begin{figure}
    \centering
    \includegraphics[width=0.49\textwidth]{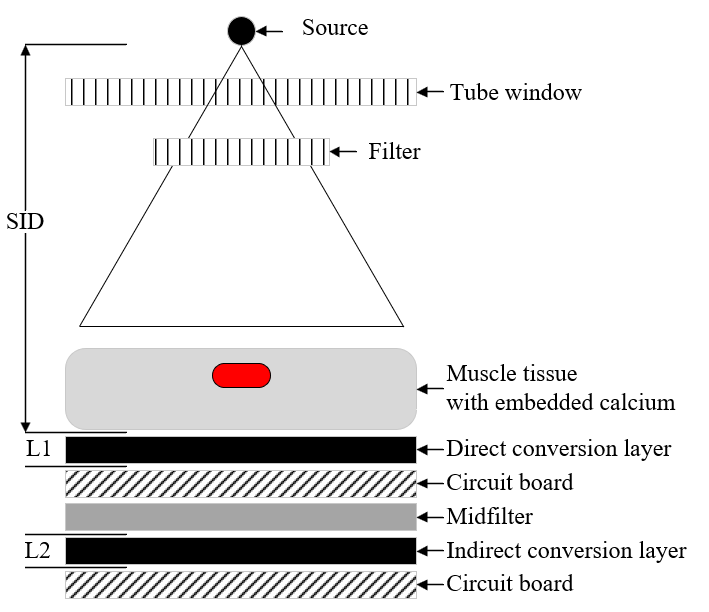}
    \caption{Illustration of the single-exposure DE imaging system simulated using the CLSM. The detector architecture involves an a-Se top detector, a 1 mm Cu interdetector filter, and a CsI:Tl bottom detector. The circuit board layers are not considered in the CLSM.}
    \label{geomtry}
\end{figure}

For the purpose of the simulation, the incident X-ray energy spectrum was generated using Spektr 3.0, based on data from the TASMIP method \cite{punnoose2016spektr}. The simulations in this paper were performed using a polyenergetic 120 kVp X-ray spectrum with a source to image distance (SID) of  100 cm.  
The spectrum is first absorbed by soft tissue with a 1 mm embedded calcification (density of 1200 mg/cm$^{3}$ \cite{kirkbride2017discrimination}). A tissue thickness of 22.6 cm was determined as an average from three simulated patient thicknesses, which were later used in the determination of anatomical noise parameters. The spectrum then passes through the top a-Se detector with variable thickness,
a 1 mm Cu interdetector filter, then to the bottom detector CsI:Tl with a thickness of 550 \textmu m.
As the top layer thickness increases, the spectrum experiences higher filtration, impacting the bottom layer. An illustration of the geometry can be seen in Fig. \ref{geomtry}.

The pixel size is 85 \textmu m for the top and bottom detectors, and the additive electronic noise (Ne\textsuperscript{-}) for each detector is assumed to be 2000 e\textsuperscript{-} per pixel. 

The following transport and operating parameters were used for the a-Se detector: K-edge energy ($E_k$) = 12.7 keV, electron mobility ($\mu_e$) = 0.003 cm$^2$/V$\cdot$s, hole mobility ($\mu_h$) = 0.12 cm$^2$/V$\cdot$s, electric field ($F$) = 10 V/\textmu m, and EHP creation energy (\emph{$W$}) based on the values from \cite{kabir2002dqe,yunus2005monte}. 

To model the CsI:Tl detector, the K-fluorescence contributions of Cs and I were computed separately and their contributions combined: $E_k$ = 35.9 and 33.2 keV respectively, and \emph{W} = 16 eV based on the values from \cite{castelli1995performance,zhao2004x}. The characteristic optical blur of the CsI:Tl scintillator was modeled using the Lorentzian fit described in \cite{cao2018modeling}. 

Using the simulated performance of the top and bottom detectors, the theoretical framework developed by Richard et al.~\cite{richard2005generalized, richard2007optimization} is employed to understand and optimize the DE imaging performance of the DL detector. The detectability index ($d'$) provides a figure of merit that incorporates detector performance and imaging task, and offers an objective function for system optimization:
\vspace{0.25cm}
\begin{equation}
d'^{2}= \displaystyle \int_{-u_{Nyq}}^{u_{Nyq}} \int_{-v_{Nyq}}^{v_{Nyq}} \text{GNEQ}^{\text{DE}}(u,v)W_{\text{task}}(u,v) du\, dv.
\label{Detectability}
\end{equation}
\vspace{0.25cm}

The spatial frequency content of the task function,
$F_{\text{task}}(u,v)$, is determined by the Fourier transform of the difference between the object function and the background
function,
\begin{equation}
F_{\text{task}}(u,v) = \displaystyle N |\mathcal{F}(H_{1}(x,y) - H_{2}(x,y))|
\end{equation} 
where $H_{1}(x,y)$ and $H_{2}(x,y)$ are hypotheses corresponding to the object and background functions \cite{vennart1997icru}. Although several tasks are relevant to chest imaging (e.g., discrimination, size estimation, and localization), a 1 mm spherical 
detection task ($W_{\text{task}}(u,v)$) is applied for identification of CAC \cite{burgess2001human,fredenberg2010observer}. The imaging task was defined directly in the Fourier domain to investigate the detectability index  ($d'$):

\begin{equation}
W_{\text{task}}(u,v) = \displaystyle \Delta I\times F_{\text{task}}(u,v).
\end{equation}

The contrast factor ($\Delta I$), where $I^{\text{obj}}$ is the signal behind the object (calcification) and $I^{\text{back}}$ is the signal behind the background, was derived with the following expression:

\begin{equation}
\Delta I = -\left( \left( \frac{I_\text{H}^{\text{obj}}}{\langle I_\text{H}^{\text{back}} \rangle} - 1 \right) - w_s \left( \frac{I_\text{L}^{\text{obj}}}{\langle I_\text{L}^{\text{back}} \rangle} - 1 \right) \right)
\label{Delta-I}
\end{equation}
where subscripts $L$ and $H$ stand for LE and HE, respectively.

The tissue cancellation parameter ($w_s$) is a user-defined weighting factor that controls the relative contribution of the LE and HE detectors (e.g., setting $w_s$ = 0 results in an image based solely on the signal from the bottom detector, while higher values of $w_s$ increase the contribution of the top detector). $w_s$ was evaluated over a range from 0 to 1.4 to assess the performance dependence between the top and bottom layer. To quantify the contrast ($\Delta I$) within the task function ($W_{\text{task}}$), we consider the LE and HE projections through the object as $I_{\text{L,H}}^{\text{obj}}$, and the average background signal as $<I_{\text{L,H}}^{\text{back}}>$.

The MTF is used to evaluate the spatial resolution characteristics of each detector and the combined performance as a DE imaging system. The MTF$_{\text{DE}}$ is derived below for the general case to account for any potential differences in the MTF in the LE and HE images\cite{allec2011analysis}:
\begin{equation}
\text{MTF}_{\text{DE}}(u,v) = \displaystyle \left|\frac{-w_{s} w_{\text{L}} \text{MTF}_{\text{L}}(u,v) + w_{\text{H}} \text{MTF}_{\text{H}}(u,v)}{-w_{s} w_{\text{L}} + w_{\text{H}}}\right|
\end{equation}
where $w_\text{L} = I'^{\text{obj}}_{\text{L}}-I'^{\text{back}}_{\text{L}}$ and $w_\text{H}= I'^{\text{obj}}_{\text{H}}-I'^{\text{back}}_{\text{H}}$ are weights associated with the zero mean signal, $I'$, in the LE and HE projections after the object (obj) and background (back), respectively.

There are two primary contributions to NNPS: that associated with the dual-energy system, NNPS$_{\text{DE}}$, and that from anatomical noise, NNPS$_\text{B}$. NNPS$_\text{DE}$ is the resulting noise from the weighted contribution of each detector. It was previously determined that the quantum noise in the top and bottom layers of a multilayer detector are uncorrelated and independent \cite{sones1989noise}, and is represented as 
\begin{equation}
\text{NNPS}_{\text{DE}}(u,v) = {\text{NNPS}_\text{H}(u,v) + w_{s}^{2}\text{NNPS}_{\text{L}}(u,v)}
\label{NNSPDE-ws}
\end{equation}
where NNPS$_{\text{L,H}}$ is the output normalized noise power spectrum from the CLSM.

Anatomical noise, NNPS$_\text{B}$, is defined as 
\begin{equation}
    \text{NNPS}_\text{B} (f) = \left\langle \left. \frac{dI}{dg} \right|^2 \right\rangle \frac{K}{f^\beta}
\label{k-beta}
\end{equation}
where $f$ is frequency, $I$ is the detected signal, $dg$ is the variation in the tissue, and K and $\beta$ are anatomical noise parameters. 

For the purposes of this work, NNPS$_{\text{B}}$ was derived using an average of three simulated DE lateral-view patient chest images \cite{richard2005generalized}, with an average thickness of 22.6 cm; the generation of these images is detailed in Section \ref{image simulation}. Using a weighted logarithmic subtraction (WLS), the LE and HE images were combined to yield a single anatomical image:
\begin{equation}
    I_{\text{DE}}(x,y) = \ln(I_\text{H}(x,y)) - w_{s} \ln(I_\text{L}(x,y)).
    \label{IDE}
\end{equation}

From this single image, the region of interest (ROI) around the heart was manually delineated and divided into nonoverlapping realizations. 
The initial image was 5060 $\times$ 5060 pixels, with the ROI defined as an area of 360 $\times$ 360 pixels, and the size of each realization was 180 $\times$ 180 pixels, resulting in a total of $M=4$ realizations. For each weighting factor used to create $I_{\text{DE}}$, NNPS$_\text{B}$ was calculated as

\begin{multline}
\text{NNPS$_{\text{B}}$}(u, v) = \\\\
\frac{\Delta x \Delta y}{M \cdot 180 \cdot 180}
\sum_{\text{ROI}=1}^{M} \left| \text{DFT}_{\text{2D}} \left\{ \frac{I_{\text{ROI}}(x, y) - \overline{I}_{\text{ROI}}}{\overline{I}_{\text{ROI}}} \right\} \right|^2.
\label{NNPS_SHI}
\end{multline}

\begin{figure}
\centering
\includegraphics[width=0.49\textwidth]{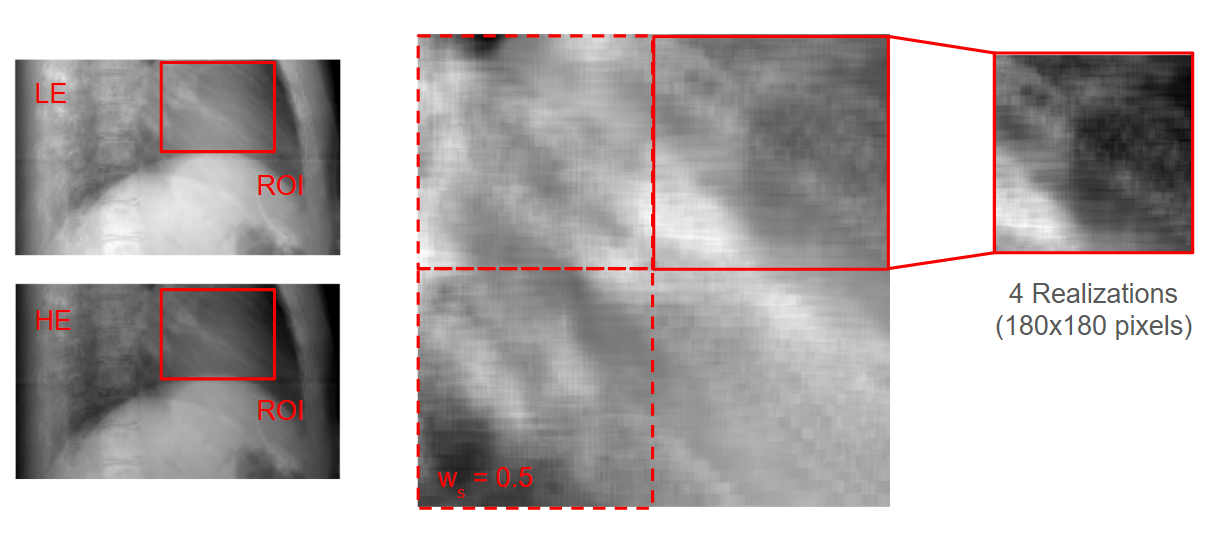}
\caption{Example lateral chest images simulated based on the LE and HE spectral separation of the single-exposure DL detector. The ROI specified corresponds to the location of the heart (and coronary arteries). After WLS (Eq.~\ref{IDE}) with $w_{s}$ = 0.5, the highlighted regions correspond to separate realizations from which, after summation, the 1D anatomical NNPS computation was derived (Eq.~\ref{NNPS_SHI}).}
\label{Anatomical process}
\end{figure}

The resulting noise curves were fit with Eq. \ref{k-beta}, providing values of $K$ and $\beta$ as a function of $w_s$ \cite{burgess2001human}.
Figure \ref{Anatomical process} shows the images used in the realization process for this derivation with $w_s=0.5$.  

Incorporating descriptions of anatomical noise in the NEQ, as a spatial-frequency-dependent noise term, NNPS$_{\text{B}}(u,v)$, included in the denominator, yields the generalized DE NEQ:

\begin{multline}
\text{GNEQ}^{\text{DE}}(u,v) = \\\\
\displaystyle \frac{\text{MTF}_{\text{DE}}^2(u,v)}{\text{NNPS}_{\text{DE}}(u,v)+\text{NNPS}_{\text{B}}(u,v)\cdot \text{MTF}_{\text{DE}}^2(u,v)}
    \label{GNEQ}
\end{multline}

Figure 4 depicts a flowchart to compute the detectability index. First, initial detector parameters, including pixel size and layer thicknesses, are selected. Second, the performance of the top and bottom detectors is simulated in the CLSM to be combined as DE contributions. Third, considering the calcification properties, $\Delta I$ and $F_{\text{task}}(u,v)$ were computed to yield the task function,
$W_{\text{task}}(u,v)$. Fourth, NNPS$_{\text{B}}(u,v)$ and NNPS$_{\text{DE}}(u,v)$ were computed to yield the GNEQ$^{\text{DE}}(u,v)$. Finally, GNEQ$^{\text{DE}}(u,v)$ and $W_{\text{task}}(u,v)$ were integrated (Eq.~\ref{Detectability}) to generate ($d'$). The detectability index was calculated over the range of $w_s$ and top layer thickness, and the peak detectability was identified.

\begin{figure*}
\centering
\includegraphics[width=0.8\textwidth]{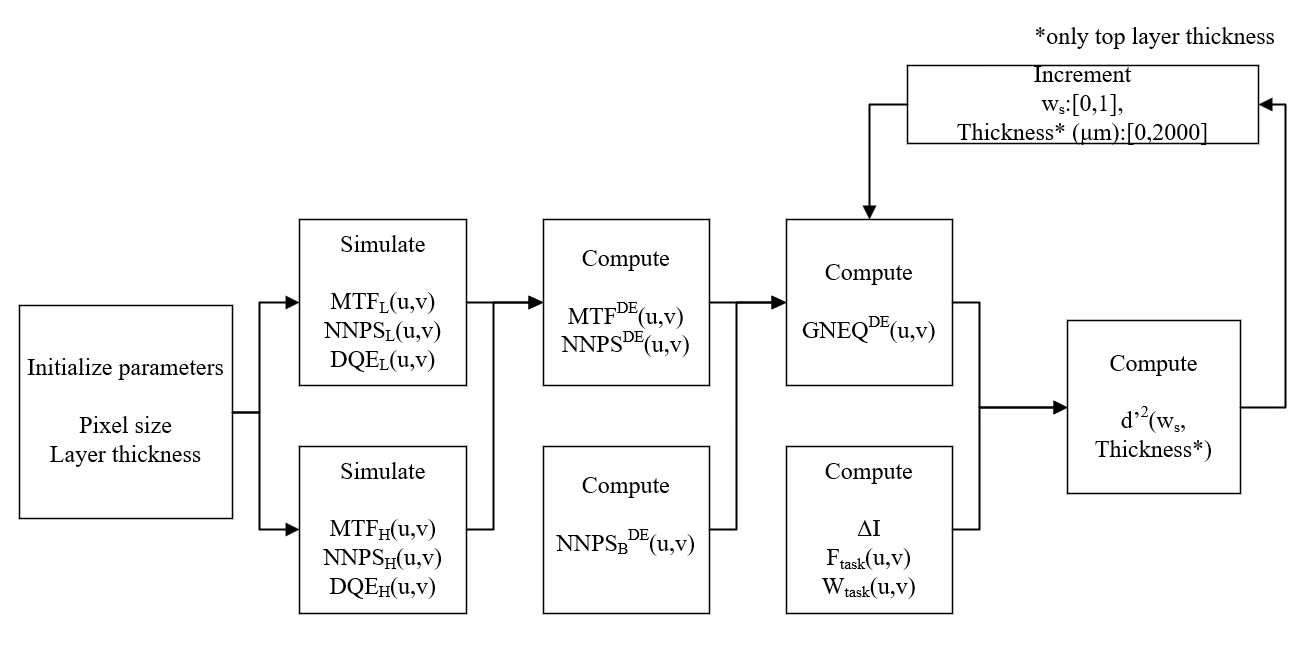}
\label{Flowchart}
\caption{Flowchart for calculation of detectability index as a function of the tissue cancellation parameter (w$_{s}$) and thickness of the top layer.}
\end{figure*}

\subsection{Image Simulation}\label{image simulation}

DE-CXR images were simulated using five CT datasets. A lateral view was chosen for imaging orientation as it has been shown to give higher CAC visibility \cite{boswell2025coronary}. Each CT had a 50-cm axial field of view (FOV) and a slice thickness of 0.625 mm, ensuring full-body coverage without truncation and supporting realistic DE-CXR simulation. For the CAC detection task, a 1 mm-diameter CAC (1200 mg/cm$^{3}$ hydroxyapatite (HA) \cite{kirkbride2017discrimination}) was inserted into the coronary region of each patient dataset.

The first step was to conduct material decomposition on the full-size CT data, which was obtained using a single energy source. The decomposition was performed based on the methods proposed by Wang et al \cite{wang2018acuros}. The resulting density maps represent the basis materials of water and calcium, which were then forward-projected onto a dual-layer flat-panel detector to generate low- and high-energy images. 

We assumed a geometry of DE-CXR with a source-detector distance of 180 cm and a source-isocenter distance of 163.5 cm. The dual-layer detector was simulated as 43 cm $\times$ 43 cm with a pixel size of 150 \textmu m. A polyenergetic spectrum from a 120 kVp source with 1 mm internal Al filtration was attenuated by the patient and then absorbed in a DL detector. The DL detector comprised a top a-Se layer with thickness varying from 20 \textmu m to 2000 \textmu m in 100-\textmu m increments, followed by a 1-mm Cu interlayer filter, and a bottom CsI:Tl scintillator layer with a thickness of 550 \textmu m. The top and bottom projections, $I_{\text{L,init}}$ and $I_{\text{H,init}}$, acquired by an energy-integrating detector can be expressed as:
\begin{equation}
    I_{\text{L,init}} = \\\\
    \int E \cdot I_{0,\text{L}}(E)\, e^{-t_w \mu_w(E) - t_b \mu_b(E)} dE,
    \label{I_top-init}
\end{equation}
\begin{equation}
    I_{\text{H,init}} = \int \! E \cdot I_{0,\text{H}}(E)\, e^{-t_w \mu_w(E) - t_b \mu_b(E)} dE
    \label{I_bot-init}
\end{equation}
where $I_{0,L}(E)$ and $I_{0,H}(E)$ are the entrance spectrum for top and bottom layers as a function of $E$ and $t_{w}$, and $\mu_{w}$ are the thickness and attenuation coefficient for basis material water while $t_{b}$ and $\mu_{b}$ are that for basis material bone. 

The second step is to inject realistic noise into the projections. To accurately model system resolution and noise correlations, we required the detector MTF and NNPS, which were taken from the cascaded-model results described in the previous section. Let \(K(\cdot)\) be the noise-shaping filter in the frequency domain, which relates to the calculated NNPS by $K(u,v) \propto \sqrt{\mathrm{NNPS}(u,v)}$, 
where \emph{u} and \emph{v} are the frequency coordinates along the \emph{x} and \emph{y} directions. The filter \(K(u,v)\) is normalized to unity so that the magnitude of the injected variance is unchanged even after the noise correlations are introduced. Let $\mathcal{F}$$\{$$\cdot$$\}$ be the 2D Fourier transform and $\mathcal{F}^{-1}$$\{${$\cdot$}$\}$ its inverse. We then define: 

\begin{equation}
    \tilde{I}_{\text{L,init}}(u,v) = \mathcal{F}\left\{ I_{\text{L,init}}(x,y) \right\},
    \label{aveI_top-init}
\end{equation}
\begin{equation}
    \tilde{I}_{\text{H,init}}(u,v) = \mathcal{F}\left\{ I_{\text{H,init}}(x,y) \right\}, \label{aveI_bot-init}
\end{equation}
\begin{equation}
    \tilde{N}_{\text{L}}(u,v) = \mathcal{F}\left\{ \sigma\!\left(I_{\text{L,init}}(x,y)\right) \cdot n_L(x,y) \right\},
    \label{N_top}
\end{equation}
\begin{equation}
    \label{N_bot}
    \tilde{N}_{\text{H}}(u,v) = \mathcal{F}\left\{ \sigma\bigl(I_{\text{H,init}}(x,y)\bigr) \cdot n_H(x,y) \right\},
\end{equation}
where the standard deviation map $\sigma$({$\cdot$}) is computed as

\begin{equation}
    \sigma (I_{\text{L,init}}) = \sqrt{\int E^2 \cdot I_{0,\text{L}}(E)e^{-t_w \mu _w(E)-t_b \mu _b(E)}\, dE}\,\text{,}
\end{equation}
\begin{equation}
    \sigma (I_{\text{H,init}}) = \sqrt{\int E^2 \cdot I_{0,\text{H}}(E)e^{-t_w \mu _w(E)-t_b \mu _b(E)}\, dE}\,\text{.}
\end{equation}

Multiplying $\sigma$({$\cdot$}) with white noise \(n_L(x,y),n_H(x,y) \sim \mathcal{N}(0, 1)\) generates uncorrelated noise with the desired magnitude to inject, and the degree of correlation is introduced using filter \(K(u,v)\). Thus, the noise injection process can be expressed as:

\begin{multline}
    I_{\text{L}}(x,y) = \\\\
    \mathcal{F}^{-1} \left\{ \tilde{I}_{\text{L,init}}(u,v)\,\mathrm{MTF}_{\text{L}}(u,v) + \tilde{N}_{\text{L}}(u,v)\,K_{\text{L}}(u,v) \right\}
    \label{I_top}
\end{multline}
\begin{multline}
    I_{\text{H}}(x,y) = \\\\
    \mathcal{F}^{-1} \left\{ \tilde{I}_{\text{H,init}}(u,v)\,\mathrm{MTF}_{\text{H}}(u,v) + \tilde{N}_{\text{H}}(u,v)\,K_{\text{H}}(u,v) \right\}
    \label{I_bot}
\end{multline}
$I_{\text{L}}$ and $I_{\text{H}}$ are then log normalized to convert to line integrals $l_{\text{L,H}} = -ln(I_{\text{H,L}}/I_0)$. The weighted subtraction between $l_{\text{L}}$ and $l_{\text{H}}$ then generates a bone image, which is used to optimize CAC visibility as 
\begin{equation}
    l_{\mathrm{CAC}} = - (l_{\mathrm{H}} - w_s \cdot l_{\mathrm{L}}) .
    \label{l_CAC}
\end{equation}

We simulated images at 50 mAs, matching the exposure used in the cascaded model study. 

\subsection{CNR Simulation}
CAC visibility was evaluated as a function of a-Se thickness and the weighting factor $w_s$, and was quantified using the CNR computed between the known CAC region on and a relatively uniform background (outside the bone region) as
\begin{equation}
    \text{CNR} = l_{\text{CAC}}/\sigma_{Bg}.
\end{equation} 
The CNR was then averaged across the five simulated patients to obtain the final result.

\section{Results \& Discussion}

\begin{figure}
\centering
\includegraphics[width=0.49\textwidth]{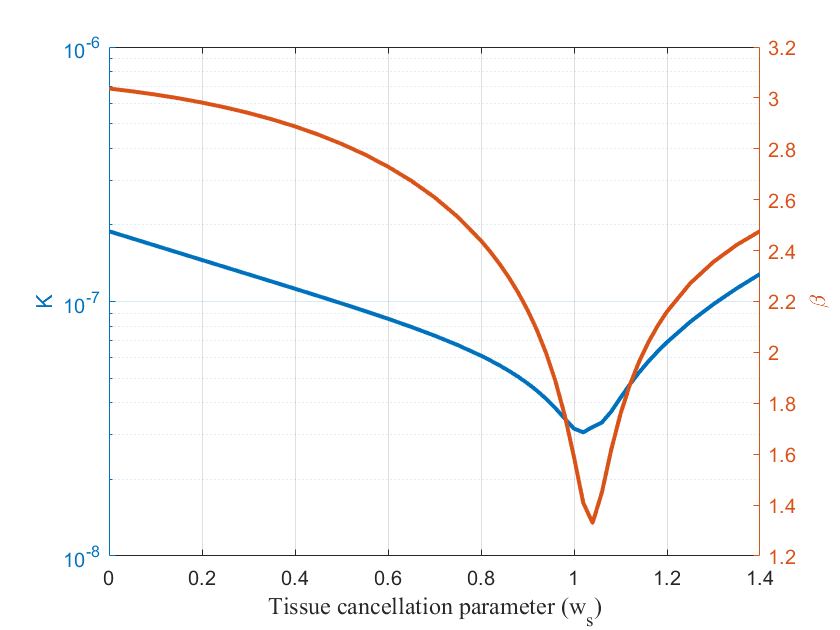}
\caption{Anatomical noise parameters, K and $\beta$, measured as a function $w_{s}$ showing a dependence on the cancellation of the combined DE image.}
\label{Anatomical}
\end{figure}

A central factor in the CLSM analysis was the treatment of anatomical noise. For small, high-contrast objects such as CAC, visualization in chest radiography is often limited not only by quantum noise but also by structured anatomical background from overlapping ribs, spine, soft tissue, and cardiac anatomy. The tissue cancellation parameter $w_s$ determines how the low- and high-energy images are combined and therefore controls the amount of residual anatomical texture in the DE image. As shown in Fig. \ref{Anatomical}, the fitted anatomical-noise parameters $K$ and $\beta$ both depend strongly on $w_s$, with minima occurring slightly above $w_s=1$. This trend indicates that approximately balanced contributions from the low- and high-energy detector layers provide the strongest suppression of anatomical structure in the heart region.

The behavior of $K$ is particularly important because it acts as a multiplicative scale factor in NNPS$_\text{B}$. Reductions in $K$ directly reduce the magnitude of anatomical noise over the spatial-frequency range included in the detectability calculation. The parameter $\beta$, by contrast, changes the slope of the anatomical-noise spectrum and therefore redistributes the residual anatomical noise across spatial frequency. Because the CAC task is localized and has substantial frequency content near the millimeter scale, changes in $\beta$ can still affect detectability, but their influence is more frequency dependent than that of $K$. The minimum in $\beta$ near the tissue-cancellation weight may also reflect that once anatomical structure is strongly suppressed, the residual spectrum contains a larger relative contribution from system noise that cannot be removed by changing $w_s$.

\begin{figure}
\centering
\includegraphics[width=0.49\textwidth]{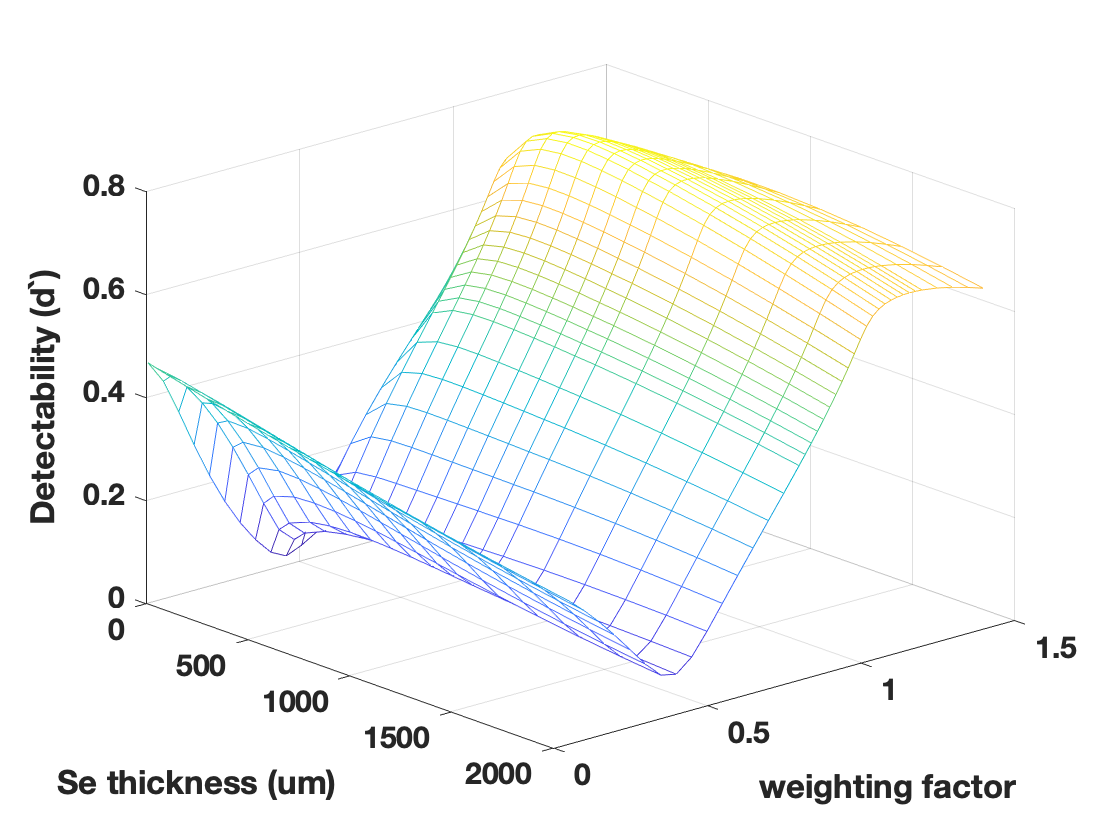}
\vspace{0.25cm}
(a)

\includegraphics[width=0.49\textwidth]{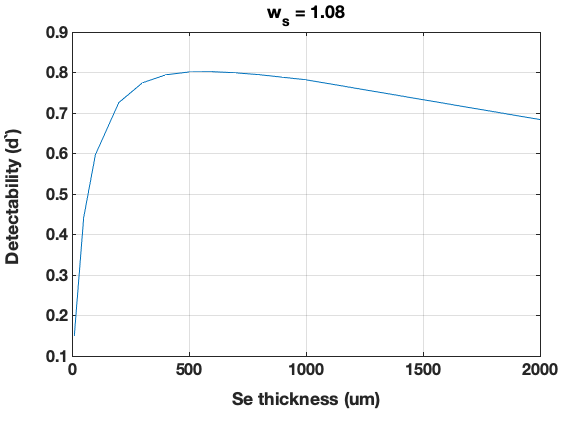}
(b)
\caption{Detectability index computed as a function of the top layer a-Se thickness (in \textmu m)  and tissue cancellation parameter (w$_s$). a) The optimal parameters of 500-550 \textmu m top layer a-Se thickness with $w_{s}$ = 1.08 provide peak detectability, $d'$. b) Detectability, $d'$, as a function of the top layer a-Se thickness at $w_{s}$ = 1.08.}
\label{Detect}
\end{figure}

After incorporating the fitted anatomical-noise parameters into the GNEQ calculation, detectability was evaluated as a function of both a-Se top-layer thickness and $w_s$. Figure \ref{Detect}a shows that the maximum CLSM detectability occurs near $w_s=1.08$ and an a-Se thickness of approximately 500–550 \textmu m. The dependence on $w_s$ is stronger than would be expected from detector noise alone, emphasizing that anatomical-noise cancellation is a dominant factor in this task. The pronounced reduction in detectability near $w_s \approx 0.5$ is likely caused by a combination of reduced calcification task contrast (Eq. \ref{Delta-I}) and increased weighted contribution of top-layer noise (Eq. \ref{NNSPDE-ws}), without sufficient suppression of anatomical background. In this region, the low- and high-energy CAC signals are partially cancelled in the weighted subtraction, but the residual anatomy has not yet reached its minimum. As $w_s$ approaches the tissue-cancellation region near unity, anatomical noise decreases, and detectability recovers.

Although the peak CLSM detectability occurs near 500–550 \textmu m, Fig. \ref{Detect}b shows that performance is relatively flat over a broad range of a-Se thicknesses. At $w_s=1.08$, thicknesses from roughly 300 to 1000 \textmu m yield detectability values within a few percent of the maximum. This broad optimum is important from a detector-design perspective. Increasing a-Se thickness improves low-energy absorption and spectral separation up to a point, but very thick a-Se layers reduce photon fluence reaching the bottom CsI:Tl detector and increase the high voltage required to maintain the desired electric field \cite{kasap2002direct}. Therefore, a thinner layer such as 300 \textmu m may provide a more practical design choice, preserving near-optimal CAC detectability while reducing fabrication and operation challenges associated with thicker a-Se layers \cite{huang2019recent, hellier2024evaluating}. As the CLSM response is broad across this range, the exact numerical optimum is less important than identifying a practical operating region in which detectability remains high while manufacturability, required bias voltage, charge-collection stability, and long-term detector reliability remain feasible.

\begin{figure}
\centering
\includegraphics[width=0.49\textwidth]{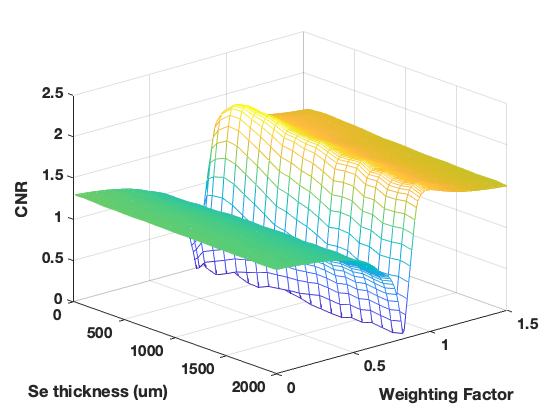}
\caption{Contrast-to-noise ratio (CNR) as a function of a-Se thickness and the weighting factor ($w_{s}$) between the top and bottom detector layers. The CNR peaks at an a-Se thickness of 300 \textmu m and $w_{s}$ = 0.93.}
\label{CNR0}
\end{figure}

The CT-based image simulations provide an additional image-domain assessment of these CLSM trends, shown in Fig. \ref{CNR0}. In the patient-based simulations, the highest average CNR occurred at an a-Se thickness of 300 \textmu m and $w_s=0.93$, close to the near-equal weighting suggested by the CLSM. Like the CLSM, a range of thicknesses provides an acceptable CNR, though this falls more narrowly across 200-500 \textmu m. The lower optimal thickness in the simulated image study is reasonable because this simulation includes realistic anatomical heterogeneity, spatially varying attenuation, projection overlap, and ROI-based CNR measurement, whereas the CLSM uses a simplified uniform-thickness object and an ideal-observer detectability metric. Thus, the exact numerical optima are not expected to be identical. More importantly, both studies support the same design conclusion: CAC conspicuity is maximized when the two detector layers contribute comparably to the DE subtraction and when the a-Se layer is thick enough to provide useful low-energy information but not so thick that it substantially compromises the bottom-layer signal.

\begin{figure*}
\centering
\includegraphics[width=\textwidth]{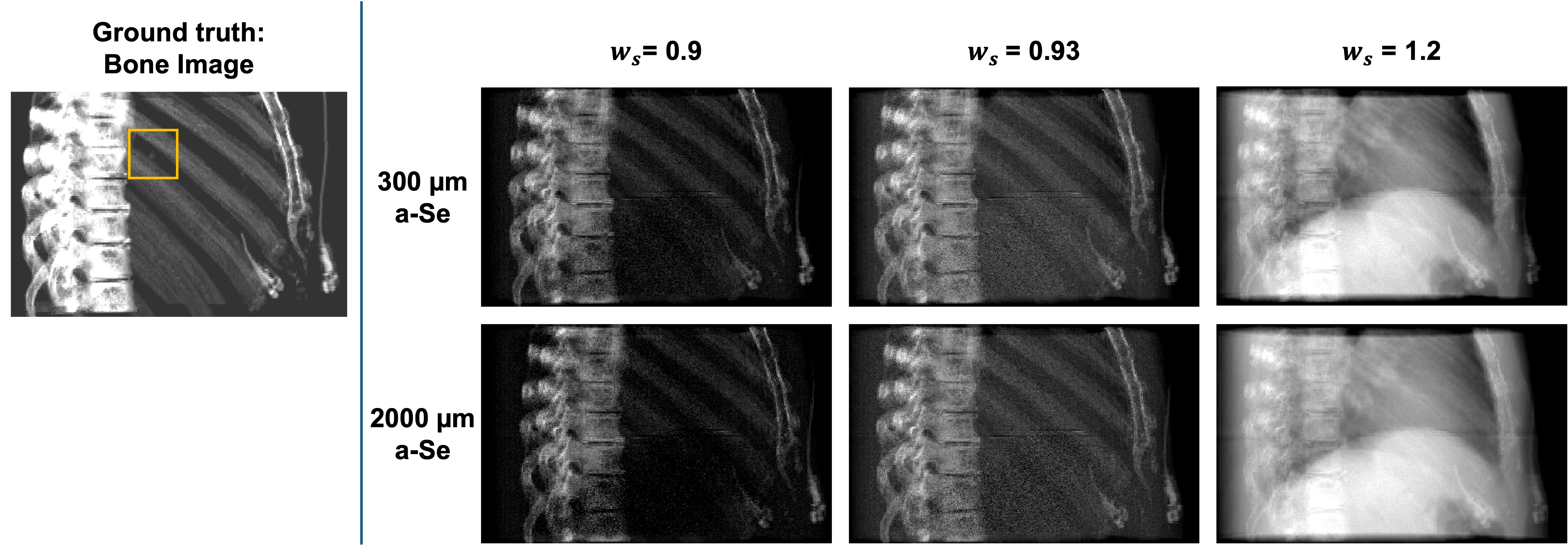}

(a)

\includegraphics[width=\textwidth]{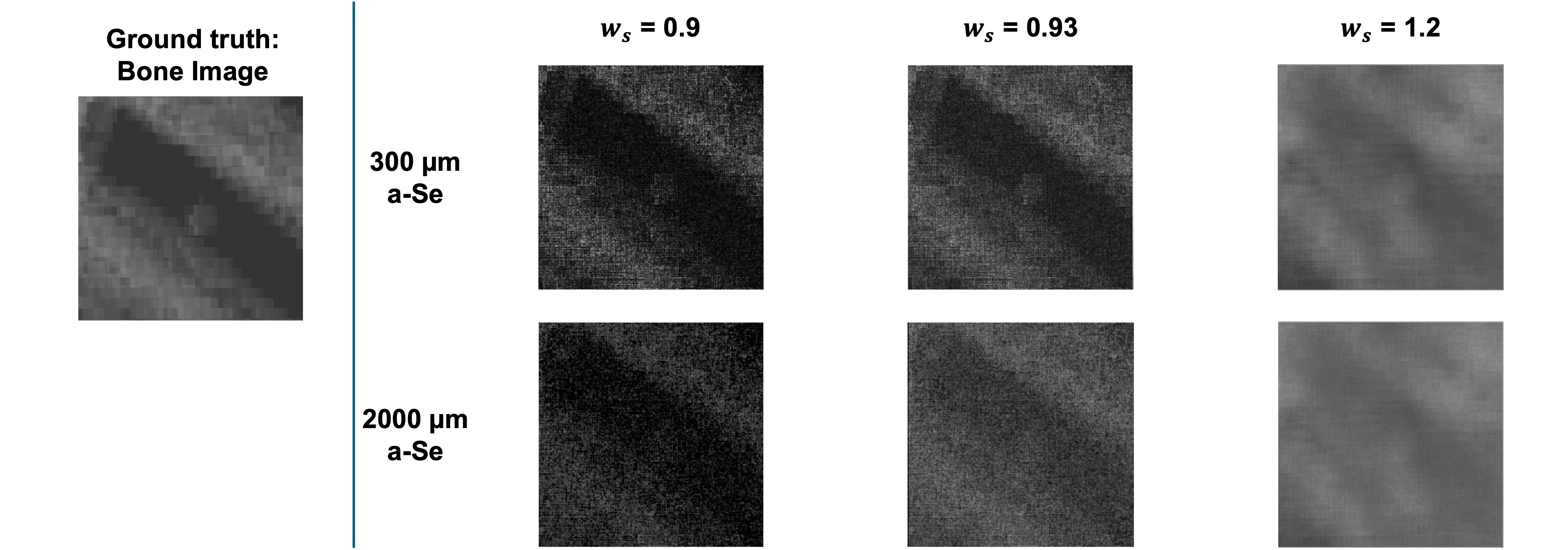}

(b)

\caption{Impact of a-Se thickness and weighting factor $w_s$ on CAC conspicuity in simulated DE-CXR. All images utilize the same grayscale window/level.
a) Representative $l_{\text{CAC}}$ generated using different combinations of a-Se top-layer thickness and weighting factor $w_{s}$. The yellow box indicates the region containing the inserted CAC.
b) Zoom-in of the boxed region highlighting how CAC visibility changes with a-Se thickness and $w_{s}$; the lesion is most conspicuous at the optimal setting (300 \textmu m a-Se, $w_{s}$ = 0.93).}
\label{CNR12}
\end{figure*}

The representative images in Fig. \ref{CNR12} further illustrate the importance of the weighting factor. At the optimal weighting factor, the DE-derived CAC image suppresses much of the soft-tissue background while retaining the inserted calcification signal. In comparison, nonoptimal weights either leave excess anatomical structure or reduce the conspicuity of the calcification itself. This behavior supports the interpretation of $w_s$ as a tissue-cancellation parameter rather than simply a detector-combination weight. The optimum is therefore governed by the balance between removing anatomical clutter and preserving the spectral contrast of calcium.

Overall, the consistency between the task-based detectability and image-domain CNR endpoints supports a direct a-Se/indirect CsI:Tl DL detector as a promising configuration for opportunistic CAC assessment using single-exposure DE-CXR. The CLSM identifies the physical mechanisms that drive detectability, while the simulated image study demonstrates that the same trends persist in realistic patient anatomy. Taken together, the results suggest that a 300 \textmu m a-Se top layer is a practical design candidate: it is near the broad CLSM detectability plateau, coincides with the simulated image CNR optimum, and may reduce the high-voltage and fabrication challenges associated with thicker a-Se layers. This demonstrates these models as effective ways to evaluate optimal a-Se thicknesses and top/bottom weighting parameters. Future work will continue to employ these models, investigating smaller calcifications, and limits of detection for the direct/indirect DL detector system.

\section{Conclusion}
An integrated simulation framework was used to evaluate a direct a-Se/indirect CsI:Tl dual-layer detector for opportunistic CAC assessment in single-exposure dual-energy chest radiography. By combining CLSM-derived detector performance metrics, image-derived anatomical-noise estimates, and CT-based image-domain CNR analysis, the framework assessed how a-Se top-layer thickness and dual-energy weighting affect CAC detectability and conspicuity.

Both models indicated near-equal weighting of the low- and high-energy detector layers improved anatomical-noise suppression while preserving calcium contrast. Although the CLSM detectability peak occurred at a thicker a-Se layer, the broad detectability plateau and image-domain CNR results support a 300 \textmu m a-Se top layer as a practical design candidate. This thickness provides near-optimal task performance while reducing high voltage requirements and fabrication challenges associated with thicker a-Se layers.

Future work will extend this analysis to additional detector parameters, including interdetector filter material and thickness and CsI:Tl scintillator thickness, as well as smaller calcifications and varied CAC densities and morphologies. Experimental validation using a physical detector prototype and anthropomorphic phantoms will be necessary to confirm the predicted optimal design range and assess clinical feasibility for opportunistic CAC assessment in routine chest radiography.

\section*{Acknowledgments}
This work was supported in part by the National Institutes of Health, National Institute of Biomedical Imaging and Bioengineering, award number R01EB033466.

The authors would like to thank Dr. Nicholas Allec for help in simulation and data analysis.

\bibliographystyle{IEEEtran}
\bibliography{mybib.bib}
\end{document}